\documentstyle[sprocl]{article}

\def\Journal#1#2#3#4{{#1} {\bf #2}, #3 (#4)}

\def\NIMA{{\em Nucl. Instrum. Methods} A}
\def\NPB{{\em Nucl. Phys.} B}
\def\NPA{{\em Nucl. Phys.} A}
\def\PH{{\em Physics}}
\def\PLB{{\em Phys. Lett.}  B}
\def\PRL{\em Phys. Rev. Lett.}

\def\PRC{{\em Phys. Rev.} C}

\def\RPP{{\em Rep. Prog. Phys.}}
\def\JPA{{\em J. of Phys.} A}
\def\JPG{{\em J. o Phys.} G: Nucl. Part. Phys.}
\def\JP{{\em J. de Phys.}}

\def\be{\begin{equation}}
\def\ee{\end{equation}}
\def\bea{\begin{eqnarray}}
\def\eea{\end{eqnarray}}

\begin{document}

\title{ CRITICAL BEHAVIOUR IN PERIPHERAL Au + Au COLLISIONS
AT 35 MeV/u }

\author{
M. Bruno$^1$, P. F. Mastinu$^{1,3}$, M. Belkacem$^{1,9}$, M. D'Agostino$^1$,
P. M. Milazzo$^{1,2}$, G. Vannini$^2$,
D. R. Bowman$^7$, 
J. D. Dinius$^6$,
A. Ferrero$^{4,8}$, M. L. Fiandri$^1$, C. K. Gelbke$^6$, T. Glasmacher$^6$,
F. Gramegna$^5$, D. O. Handzy$^6$, D. Horn$^7$, W. C. Hsi$^6$, M. Huang$^6$,
I. Iori$^4$, G. J. Kunde$^6$,
M. A. Lisa$^6$, W. G. Lynch$^6$,
G. V. Margagliotti$^2$, C. P. Montoya$^6$, A. Moroni$^4$, G. F. Peaslee$^6$,
R. Rui$^2$,
C. Schwarz$^6$, M. B. Tsang$^6$, C. Williams$^6$,
V. Latora$^9$ and A. Bonasera$^9$
}

\address{
$^{1}$ Dipartimento di Fisica and INFN, Bologna, Italy \\
$^{2}$ Dipartimento di Fisica and INFN, Trieste, Italy \\
$^{3}$ Dipartimento di Fisica, Padova, Italy \\
$^{4}$ Dipartimento di Fisica and INFN, Milano, Italy \\
$^{5}$ INFN, Laboratori Nazionali di Legnaro, Italy \\
$^{6}$ NSCL, Michigan State University, USA \\
$^{7}$ Chalk River Laboratories, Chalk River, Canada \\
$^{8}$ On leave from CNEA, Buenos Aires, Argentina \\
$^{9}$ INFN, laboratorio Nazionale del Sud, Catania, Italy
}


\maketitle\abstracts{
The signals theoretically predicted for the occurrence
of a critical behavior (conditional moments of
charge distributions, Campi scatter plot, fluctuations of the size
of the largest fragment, power law in the charge distribution, intermittency) 
have been found for peripheral events in the reaction Au+Au at 35 MeV/u.
The same signals have been studied with
a dynamical model which foresees phase transition, like the Classical 
Molecular Dynamics. }
  
\section{Introduction}
The liquid-gas phase transition in nuclear systems
has been recently theoretically and experimentally 
investigated~\cite{various,eos,belkacem}. 
In this contribution we report 
on the search for critical behavior signals in the
experimental data of the Au + Au reaction at 35 MeV/u.
The experiment was performed at the National Superconducting
Cyclotron Laboratory of the Michigan State University taking advantage of 
the coupling of the Multics~\cite{exp1}  and 
the Miniball~\cite{exp2} apparatus. The analysis has been restricted
to events corresponding 
to peripheral and semi-peripheral reactions, selected imposing that
the component of the velocity of the largest fragment along the beam 
direction was greater
than 75\% of the beam velocity and that the total detected charge was between
70 and 90. Most of these events come from the disassembly
of the quasi-projectile, since the detection of the quasi-target is
suppressed due to the thresholds of the detectors.

Comparisons with the results of calculations in the
framework of Classical Molecular Dynamics model
(CMD)~\cite{belkacem,CMD} are also presented.

\section{ Experimental results}

The events selected with the previously mentioned criterion have been plotted
(see Fig.~1) as $ln(Z^j_{max})$ vs. $ln(M^j_2)$ 
(Campi scatter plot~\cite{campi}), where $Z^{j}_{max}$
is the charge of the heaviest fragment and $M^{j}_{2}$
is the second conditional moment of the charge distribution
detected in the $j$-th event,
$M^{(j)}_{2} = \sum_{Z} Z^{2} n_{j}(Z) /Z_{0}$
where $n_{j}(Z)$ is the number of fragments of charge $Z$
detected in the $j$-th event, $Z_0$ is the total charge 
and the summation is over all
fragments but the heaviest detected one. 

\begin{figure}
\vskip 4cm
\caption{Campi scatter plot, $ln(Z_{max})$ versus $ln(M_{2})$.
The three different regions are discussed in the text. Fission events are 
to the right of region 2.} 
\label{cplo}
\end{figure}

The data in Fig. 1 are distributed along two branches as 
predicted by percolation calculation 
for undercritical and overcritical
events. 
To have a deeper insight in the behavior of events falling in different regions
of the Campi-plot, we selected three different cuts
in the upper branch (cut 1), in the lower one (cut
3) and in the intersection region (cut 2). 
The charged particle multiplicity
distributions observed for these three cuts show that 
cuts 1 and 3 select
low and high multiplicity events with a narrow range, whereas  for the cut 2
a wide range of charged particle multiplicities has been observed~\cite{fra},
possibly 
related to the occurrence of large fluctuations as expected at the critical
point. Even when the width of region 2 is reduced, the 
multiplicity distribution remains quite broad.

The fragment charge distribution corresponding to
cut 1 contains light fragments and heavy
residues, thus exhibiting a "U"-shaped distribution.
For cut 3 it is rapidly decreasing.
The results obtained for cut 1 and cut 3 are similar to the predictions
of percolation calculations in the sub-critical
region~\cite{bauer}, with the probability p higher than the critical 
one~(p$_C$),
and in the overcritical region (p$<$p$_C$), respectively.
Similar trends have been observed also for
the predictions from dynamical~\cite{Mohabor} and statistical 
calculations~\cite{gross1} for subcritical (T$<$T$_C$) and overcritical 
events (T$>$T$_C$), respectively.
The fragment charge distribution for cut 2 shows
a power-law distribution, $P(Z) \propto Z^{-\tau}$, 
with $\tau \approx 2.2$.
For macroscopic systems exhibiting a liquid-gas phase
transition, such a power-law distribution is predicted to
occur near the critical point~\cite{fisher}. 

An analysis in term of Scaled 
Factorial
Moments (SFM) has been performed. The SFM are defined~\cite{plocia}  as
\begin{equation}
F_i(\delta s)={{\sum _{k=1}^{Z_{tot}/ \delta s}<{n_k}\cdot ({n_k}-1)\cdot ...
\cdot({n_k}-i+1)>}
\over {\sum _{k=1}^{Z_{tot}/ \delta s}<n_k>^i}}
\label{SFM}
\end{equation}
where $Z_{tot} = 158$, and $i$ is
the order of the moment. The total interval $[1, Z_{tot}]$ is
divided into $M = Z_{tot}/\delta s$ bins of size $\delta s$, $n_k$ is the
number
of particles in the $k$-th bin for an event, and the brackets $< >$
denote the average over many events. 
The values $ln(F_{i})$ for $i$ = 2, ..., 5 are always negative (i.e. 
the variances are smaller
than for a Poissonian distribution) and almost independent on $\delta s$
for cut 3.
For cut 2, $ln(F_{i})$ are positive and almost linearly increasing as a
function of $-ln(\delta s)$ (i.e.
$F_{i} \propto \delta s^{-\lambda_{i}}$), and this,
as pointed out by several theoretical studies~\cite{belkacem,plocia},
indicates an
intermittent pattern of fluctuations~\cite{gross1,bialas,plocia}. 
Region 1, corresponding to evaporation, gives zero slope.
Increasing or reducing the sizes of the three cuts 
does not change significantly these results. 
\begin{figure}
\vskip 4cm
\caption{Relative variance $\gamma_2$ (left panel) and normalized variance of
the charge of the largest fragment $\sigma_{NV}$ (right panel) as a function of
charged particle multiplicity} 
\label{sig}
\end{figure}

Further signals which could reveal the presence of a critical behaviour have
been investigated. The second moment $M_{2}$ shows a peak versus the 
multiplicity of charged particles in the region of $N_{c} \simeq 20$.
A similar value is obtained by the EOS Collaboration~\cite{eos}.
Also the relative variance $\gamma_{2}$, defined as \cite{campi,plocia}:
$\gamma_{2} = \frac{M_{2}M_{0}}{M_{1}^{2}}$
shows a peak for 
$N_{c} \approx 18-22$ (see Fig.~2), 
consistent with that observed for $M_{2}$;
this means that around $N_{c} \approx 20$ the fluctuations in the fragment size
distributions are large, as it should be near the critical point~\cite{campi}.

A further signal for criticality, recently proposed~\cite{new}, is 
the normalized variance of the charge of the largest fragment $\sigma_{NV}$. 
This quantity, defined by:
$\sigma_{NV} = \frac{\sigma^{2}_{Z_{max}}}{<Z_{max}>}$
where
$\sigma^{2}_{Z_{max}} = <Z_{max}^{2}> - <Z_{max}>^{2}$
shows a peak at the critical point, where charge 
distributions are expected to show the largest fluctuations.
The right side of Fig.~2 shows the $\sigma_{NV}$ versus 
charged particle  multiplicity for the experimental data. 
A clear peak is present for multiplicities 
$N_{c} \approx 15 \div 20$.

The analysis of the experimental data suggests that different regions of the
nuclear phase diagram can be probed at one incident
beam energy by selecting different events~\cite{pochzo}.
All the signals so far proposed to characterize a critical behavior
have been found.
We must however caution that
the effects of finite experimental acceptance and the 
mixing of possible contributions from the decay of
projectile-like fragments and the neck-region are not yet
sufficiently well understood to allow an unambiguous
conclusion.

\section {Model calculations}
\begin{figure}
\vskip 4cm
\caption{Campi scatter plot (left panel) and normalized variance of the
charge of the largest fragment $\sigma_{NV}$ vs. charged particle multiplicity
(right panel)
for the events predicted by CMD calculations, filtered through the experimental
acceptance, with the same conditions applied to the experimental data
to select peripheral events (see text).} 
\label{sigthe}
\end{figure}
In the framework of the CMD model, calculations have been performed
for impact parameter ranging from 1 to 13 {\it fm}. 
The Campi scatter-plot for the calculated events is very
similar to the experimental one. Also M$_2$, $\gamma_{2}$ and 
$\sigma_{NV}$ show a peak for b~$\approx$~10~fm and 
for $N_{c} \approx 20-25$.
An analysis of the events in a region corresponding to the region 2 of
the experimental data show a similar behavior with a clear signal of
intermittency.

To account for the angular acceptance and detection thresholds
of the apparatus, the predictions have been suitably filtered; 
the same type of selection used to characterize the 
experimental events of peripheral or semi-peripheral reactions has 
then been applied. All the signals are clearly visible; in Fig.~3 
the Campi-plot and the $\sigma_{NV}$ with 
a peak in the region of 
b~$\approx$~10~fm and for $N_{c} \approx 15$ are shown.
The same kind of analysis performed for experimental
data gives the same results both for the charge distribution and
for the SFM.

\section{Discussion and Conclusions}

In addition to the calculations based on percolation and on statistical and 
dynamical
models, a series of simulations have been performed, mainly starting from a
given mass or charge distribution: considering the
charge and mass conservation~\cite{bao} or starting from
a simple power law mass distribution~\cite{elat}, several signals 
like the Campi plot, $M_2$, $\gamma_2$ and the intermittency have 
been observed. 
Recently a simulation based on an exponential charge 
distribution~\cite{larry} with a binomial coefficient 
shows several of the signals experimentally obtained. In particular,
peaks have been obtained for $M_2$ and $\gamma_2$,
but for instance no peak is present for $\sigma_{NV}$, as shown in Fig.~4.
The simple presence of peaks, however, could be not relevant 
and it could be important to
investigate position, height and widths of the peaks.

\begin{figure}
\vskip 4cm
\caption{Normalized variance of the
charge of the largest fragment $\sigma_{NV}$ vs. charged particle multiplicity
for the events predicted by simulations as described in the text.} 
\label{sigma}
\end{figure}

On the other side, the signals proposed so far might be not sufficient to
characterize a critical behavior or
the data extracted by the simulation
could be "somehow" connected to the experimental data which contain the 
criticality, through some constraint like the charged particle multiplicity
distribution.

In conclusion we have analyzed the peripheral events 
of the reaction Au~+~Au
at 35 MeV/u and we have found all the signals proposed to evidence out 
a critical behavior; these findings have been confirmed
by predictions based on a dynamical model like CMD.
Further work is needed before suggesting any definite conclusion.

\end{document}